\documentclass[11pt]{article}

\usepackage[margin=1in]{geometry}
\usepackage{amsmath,amssymb,amsthm,mathtools}
\usepackage{bm}
\usepackage{comment}
\usepackage{lipsum}
\usepackage[T1]{fontenc}
\usepackage[colorlinks=true,citecolor=blue,linkcolor=blue,urlcolor=blue]{hyperref}
\usepackage{subcaption}
\usepackage{xcolor}

\usepackage{enumitem}

\newtheorem{example}{Example}

\usepackage[authoryear]{natbib}

\usepackage{blindtext}
\usepackage{titling}
\usepackage{array}
\preauthor{\begin{center}
		\large \lineskip .75em%
		\begin{tabular}[t]{>{\centering\arraybackslash}p{.45\textwidth}}}
		\postauthor{\end{tabular}\par\end{center}}
\makeatletter
\renewcommand\and{%
\end{tabular}%
\hfill
\begin{tabular}[t]{>{\centering\arraybackslash}p{.45\textwidth}}}%
\makeatother

\title{The Right to be an Exception to a Data-Driven Rule}

\author{Sarah H. Cen \\ \texttt{shcen@mit.edu}\\ Massachusetts Institute of Technology \and
	Manish Raghavan \\ \texttt{mragh@mit.edu}\\ Massachusetts Institute of Technology}

\date{}

\begin{document}

\maketitle
\begin{abstract}
    Data-driven tools are increasingly used to make consequential decisions. In recent years, they have begun to advise employers on which job applicants to interview, judges on which defendants to grant bail, lenders on which homeowners to give loans, and more. In such settings, different data-driven rules result in different decisions. The problem is: to every data-driven rule, there are exceptions. While a data-driven rule may be appropriate for some individuals, it may not be appropriate for all. As data-driven decisions become more common, there are cases in which it becomes necessary to protect the individuals who, through no fault of their own, are the data-driven exceptions. 
    At the same time, however, 
    it is impossible to evaluate every one of the increasing number of data-driven decisions being made, 
     begging the question: 
     When and how should data-driven exceptions be protected?
    
    In this piece, we argue that individuals have the right to be an exception to a data-driven rule. 
    That is, the presumption should not be that a data-driven rule---even one with high accuracy---is suitable for an arbitrary decision-subject of interest. 
    Rather, a decision-maker should only apply the rule if they have exercised due care and due diligence (relative to the risk of harm) in excluding the possibility that the decision-subject is an exception to the given data-driven rule. 
    In some cases, for instance, the risk of harm may be so low that only cursory consideration is required. 
    Although applying due care and due diligence is meaningful in human-driven decision contexts, it is unclear what it means for a data-driven decision rule to do so (e.g., the ``intent'' and ``reasoning'' behind data-driven rules is often opaque). 
    We propose that determining whether a data-driven rule upholds the right to be an exception requires the consideration of three factors: \emph{individualization}, \emph{uncertainty}, and \emph{harm}.
    We unpack this right in detail, 
    providing a framework for assessing data-driven rules and describing what it would mean to invoke the right in practice.
\end{abstract}

\section{Introduction}\label{sec:intro}

We make sense of our world through rules. A bed is for sleeping while a desk is for working. Dogs have dew claws while cats have retractable ones. Doing well in school leads to good jobs.

But, to every rule, there are exceptions. Some people work on their beds, Huskies have retractable claws, and good grades do not always lead to good jobs. Although exceptions are by definition uncommon, they often carry significance disproportionate to their numbers. Exceptions not only improve our understanding of the rules, but they also help us develop better ones. Which dog breeds have retractable claws and why? If a certain dog breed has retractable claws, what other characteristics best distinguish that breed from cats?

Designing rules to distinguish between dogs and cats may be a fun thought experiment, but rules are also used to make decisions under much higher stakes. For instance, rules are used to determine which applicants a bank approves for loans, which defendants a judge grants bail, and which blobs in an image an autonomous vehicle classifies as human. In these contexts, mistakes can be harmful. An individual who fails to get a loan may lose their house, a defendant who is denied bail may be unable to provide for their dependents, and an autonomous vehicle may not stop for a person misclassified as a part of the road.

No matter how good the rule, it cannot work for everyone, begging the question: 
\emph{What happens to the individuals for which a decision rule is unfit: the exceptions?}

In some cases, nothing. We accept that rules and generalizations are, on occasion, tolerable and even necessary \citep{lippert2011we}. Indeed, the law allows landlords to put no-pet clauses in rental agreements (a rule based on the generalization that renters with pets cause more damage to homes than renters without pets) and airlines to remove passengers for safety reasons (a policy that relies on judgments about actions that a passenger could but has not yet committed). 

In other cases—typically, when the risk of harm is high—the state steps in to shield individuals from the adverse effects that can follow from the over-application of rules. Consider sentencing decisions. For many crimes, there are mandatory minimum sentences: a set of standardized rules that prescribe the minimum sentence a defendant must serve for a crime, if convicted. These rules arose in the U.S. as a way to “make sentencing procedures fairer and sentencing outcomes more predictable and consistent” \citep{travis2014growth}.
Importantly, 
mandatory minimum sentences were also used in capital cases, 
i.e., cases in which the crime is punishable by death. 
In 1976, however, the U.S. Supreme Court ruled in \emph{Woodson v. North Carolina} that mandatory minimum sentences should \emph{not} be applied to capital cases. The Court wrote that there must be “consideration of the character and record of the individual offender and the circumstances of the particular offense” before imposing a sentence as serious and irrevocable as the death penalty \citep{woodson1976-2}. In other words, the Court decided that, when it comes to the death penalty, rules that regularly yield exceptions—in this setting, defendants on which the rule, but not the presiding judge, would impose the death penalty—are unacceptable. The courts responded by giving greater discretion to judges.

\subsection*{The Right to be an Exception to a Data-Driven Rule}

In this piece, we turn our attention to \emph{data-driven rules}.
By “data-driven rules,” we refer to the decision rules behind data-driven decision aids. For example, data-driven decision aids in lending advise lenders on whether or not to grant a loan. Generally speaking, these aids produce a score for each applicant that predicts the likelihood that the applicant, if approved, would repay the loan. Different rules generate different scores. While one rule may give higher scores to applicants with families, another may not. One rule may use the applicant’s zip code as an input while another may not. As such, an applicant may be approved for a loan under some rules but not others.

As many scholars have acknowledged,
there is a gap in the governance of data-driven decisions because individuals who are subject to data-driven decisions are not always protected by a legal system that has been built around human decisions
\citep{citron2007technological,wachter2019right,kaminski2021right}.
In an effort to close this gap,
we argue that individuals have the \emph{right to be an exception to a data-driven rule}. 

This right is built on the following intuition. 
For a given context, there are many possible data-driven rules. 
Each rule yields a different set of exceptions. 
In most (if not all) cases, 
it is impossible to know who the exceptions are. 
Indeed, if a decision-maker could identify a rule's exceptions \emph{a priori}, 
then they would be able to correct the rule on those individuals,
which would imply that the decision-maker knows the correct decisions and does not need a data-driven aid to begin with.
 
Therefore, 
for every data-driven decision rule, 
it is inevitable that some individuals are---through no fault of their own---the exceptions. 
The problem is: data-driven rules do not behave like human ones. 
Add a new data point to the training set, 
and the resulting data-driven rule may completely change. 
Slightly modify the objective function and the same may occur. 
Because every data-driven rule yields a different (often unpredictable) set of exceptions, 
\emph{a decision-maker should not presume that a data-driven rule is suitable for the given decision-subject when the risk of harm is high}. 
Indeed, while the data-driven rule may work well for some individuals, there is no guarantee that it works well for the individual under consideration. 
A decision-maker should instead consider the possibility that the decision-subject is an exception to the data-driven rule,  
giving a level of consideration that is appropriate to the decision's risk of harm (in some cases, the risk of harm may be so low that no more than cursory consideration is required). 
When the risk of harm is high,
the decision-maker should presume that the decision-subject \emph{is} an exception, inflicting harm only if they are sufficiently confident that the decision-subject is {not} an exception.
 
That is, a decision-subject has the right to be an exception to the data-driven rule.
In this piece, we unpack this right in detail. 
We begin in Section \ref{sec:data_driven_exceptions} by examining why data-driven exceptions deserve special attention and how they arise. 
We then position the right to be an exception relative to existing concepts in Section \ref{sec:background}.
In Section \ref{sec:right}, 
we %
lay out a framework for determining whether a data-driven decision respects an individual's right to be an exception to a data-driven rule, 
arguing that the right requires the consideration of three factors: 
\emph{individualization}, \emph{uncertainty}, 
and \emph{harm}. 
In Section \ref{sec:operationalizing}, 
we consider what it means (and does not mean) to invoke the right to be an exception and how the right can be operationalized. 
Importantly, 
invoking the right to be an exception does not mean proving that the data-driven rule made a mistake on the decision-subject of interest (in which case it would be similar to the right to rectification), 
nor does it imply that data-driven rules should not be used at all, 
as explained in Section \ref{sec:invoking}.
We conclude in Section \ref{sec:gap} with a discussion on the potential benefits of the right to be an exception, 
touching on issues of contestation, legitimacy, transparency, and burden of proof.

\section{Data-Driven Exceptions} \label{sec:data_driven_exceptions}

Data-driven rules are behind an increasing number of consequential decisions. They inform decisions not only in lending, but also in hiring, college admissions, healthcare, social welfare, and criminal sentencing, to name a few. For instance, over 90 percent of organizations now use automated algorithms to screen and track resum\'{e}s \citep{gupta2022}. 

As the reach of data-driven rules continues to grow, we must ask: 
\emph{Who are the exceptions to each data-driven rule}? Specifically, \emph{who are the individuals that the rule fails to capture, and are there appropriate protections for these individuals}?
In this section, 
we begin by examining why data-driven exceptions deserve special attention and how they arise in the data-driven pipeline.

\subsection{Why Data-Driven Exceptions Matter}
In some sense, exceptions are exceptions, regardless of their origins. After all, both humans and machines use rules to make decisions, and all rules have blind spots. There are, however, several reasons why data-driven exceptions deserve special attention.

For one, \emph{averages appear in every part of the data-driven decision-making pipeline, making exceptions unavoidable}. For instance, the most popular data-driven methods---including maximum likelihood estimation \citep{bishop2006pattern} and empirical risk minimization \citep{devroye2013probabilistic}---use averages as their objective functions. Not only that, but the most common performance metrics---including accuracy, precision, and recall \citep{fawcett2006introduction}---are averages. High accuracy, for example, indicates that a decision rule performs well on average. The pervasive use of averages in data-driven decision-making matters because, by definition, averages work well on average, but there is no guarantee that they work well for a specific individual. When the population is heterogeneous, an average-based rule always makes mistakes. The message is not that we must abandon averages, but that data-driven assessments should be viewed for what they are \citep{lippert2011we}. Indeed, an algorithm that is known to work well in aggregate should be applied with caution when the decision-maker cares about the outcomes of individuals.

That modern machine learning is built on averages does not, on its own, motivate special attention to data-driven exceptions. After all, humans also rely on averages. Data-driven rules, by contrast, can be applied \emph{rapidly and repeatedly, resulting in systemic mistakes}. A popular hiring algorithm, for example, might underestimate the qualifications of individual X. While no algorithm is expected to be perfect, the scale at which data-driven algorithms are run means that individual X may be continually overlooked for job openings \citep{gray2019ghost}. Humans, on the other hand, make decisions more slowly and, by comparison, in a less patterned manner. As Arvind Narayanan once wrote, humans may make mistakes, but they do so in a diverse way: although $X$’s resumé may look unattractive to one employer, it  may appeal to another who hires $X$ \citep{benjamin2019race}. The harm that data-driven decisions can inflict is therefore more patterned and systemic. 

On top of all this, \emph{data-driven exceptions are highly non-intuitive}. Particularly when the stakes are high, both decision-makers and decision-subjects (the individuals on which the decisions are made) might wonder when the decision rule fails, how it fails, whether failure is systematic, whether they have the ability to influence the decision rule, whether there are paths for reparation, and so on. For instance, a defendant denied bail might wonder the grounds on which they were denied bail or whether providing more information would have changed the decision. While the reasoning behind human-driven decisions can generally be surfaced through intuition, the “reasoning” behind data-driven decisions is so opaque that there are now multiple research fields dedicated to interpreting them \citep{gilpin2018explaining,doshi2017towards}. For example, adding a data point to an algorithm’s training set can, in some cases, completely change its predictions (which would be akin to a human completely changing their world view in the span of a day) \citep{chen2017targeted,steinhardt2017certified}. In other cases, slightly perturbing the input in ways that are negligible to humans can flip a data-driven prediction \citep{goodfellow2014explaining,madry2017towards}. By analogy, this would be akin to a human delivering opposite decisions for inputs (57, 23.9, 99.1) and (57, 24, 99), where all three numbers are between 0 and 100.

For these and other reasons \citep{citron2007technological}, data-driven exceptions deserve special attention. This realization is, of course, not new. There are many works in law, philosophy, and computer science that study the downfalls of data-driven decision-making and their consequences \citep{citron2014scored,lippert2011we,benjamin2019race,alexander2020new}.  In this piece, we hone in on what it would mean to protect data-driven exceptions in practice. 

\subsection{How Data-Driven Exceptions Arise}

Before we delve into the right to be an exception, we unpack how data-driven exceptions arise. The takeaway is that, contrary to popular opinion, there are many causes of data-driven exceptions. It is not always the case that exceptions are simply the individuals who are poorly represented in the training data. (If they were, data-driven exceptions would be relatively predictable and, as a result, easy to handle.) 
We briefly describe four ways data-driven exceptions can arise. 
\begin{enumerate}
	\item \emph{Sampling bias} causes one type of sample $T$ to appear fewer times in the data than other types \citep{cuddeback2004detecting}. 
	For instance, suppose that a medical condition $T$ appears only once in a dataset collected over the general population.
	Unless an algorithm accounts for the fact that $T$ is rare in this dataset, 
	training on this dataset using an average-based approach (e.g., empirical risk minimization) can result in poor performance for patients of type $T$. 
	
	\item \emph{Model capacity} is a measure of a model's expressiveness \mbox{\citep{vapnik1994measuring,hu2021model}}.
	Exceptions can occur when a model's capacity is too low 
	to capture patterns beyond broad-strokes generalizations. 
	As an example, a neural network's capacity is determined by the number of nodes and layers. 
	If the relationship between the input variables $\mathbf{x}$ and the target variable $y$ is more complex than the expressiveness of the model, 
	then the model must make simplifications. 
	For instance, if the relationship between $\mathbf{x}$ and $y$ is quadratic, 
	then a linear function $f$ would not be able to capture the relationship between $\mathbf{x}$ and $y$ in its entirety.
	At best, $f$ may capture the approximate relationship between $\mathbf{x}$ and $y$ for a range of $\mathbf{x}$-values  but not all.
	The model can therefore perform well on some inputs but at the cost of performing poorly on others, and exceptions are cases $\mathbf{x}$ for which the model does not perform well.
	
	\item \emph{Distribution shift} arises when an algorithm is trained on samples drawn from one probability distribution, but the distribution on which the algorithm is deployed---or tested---is different \citep{koh2021wilds,perdomo2020performative}. The resulting data-driven rule learns patterns that hold in the training data do not necessarily hold in the test environment. For instance, one would not expect an algorithm that is trained on criminal justice data in the U.S. to perform well in the U.K. From the perspective of the initial training set, criminal cases in the U.K. look like exceptions. 

	\item \emph{Partial observability} captures scenarios where not all of the relevant information is observable \citep{kalman1963mathematical,kaelbling1998planning,hashimoto2018fairness}. 
	Suppose that two types of samples $T_1$ and $T_2$ exist. 
	Then, the data-driven rule typically performs well on one of the types---say, $T_1$---but not the other. 
	It does so because, unable to tell $T_1$ and $T_2$ apart, 
	the model lumps $T_2$ in with $T_1$ and treats them similarly. 
	For instance, if a computer science department bases graduate admissions purely on an applicant's undergraduate major and GPA, 
	it may not admit qualified applications who did not major in computer science but have relevant work experience after college. 
\end{enumerate}
Sampling bias, model (in)capacity, distribution shift, and partial observability illustrate that there are many ways that exceptions appear in machine learning, so much so that the field has developed a language with which to discuss them.
Data-driven exceptions are inevitable, 
and the goal of this piece is to lay out a framework for ensuring that individuals who---through no fault of their own---are exceptions do not fall through the cracks.

\section{Related Concepts}\label{sec:background}

The right to be exception requires that, instead of presuming that a data-driven rule can be applied to the decision-subject of interest, 
the decision-maker must first rule out the possibility that the decision-subject is an exception,
applying a level of consideration appropriate to the risk of harm. 

In this section, we examine whether existing principles underlie this right. 
We find that the concept of \emph{dignity} is closely related to the right to be an exception, provide an example of another right that also invokes dignity, 
and conclude with a note on related works.

\subsection{Dignity}\label{sec:dignity}

Bringing attention to exceptions that may be neglected by systems that work well for the majority has philosophical and legal grounding. 
One influential concept that emphasizes the importance of giving appropriate consideration and respect to each individual is \emph{dignity}.

Dignity is a concept that appears in international human rights law and domestic constitutions \citep{omahony2012dignity}. 
Despite being widely acknowledged as a ``foundational principle,'' 
its meaning and consequent role in law remain unclear \citep{omahony2012dignity,rao2011,glensy2011dignity}. 
It has been used---in different and, at times, conflicting ways---to justify the right to free speech \citep{cohencalifornia1971}, 
a gay couple's right to marry \citep{inremarriage}, 
a woman's freedom to choose an abortion \citep{plannedparenthood1992}, and more.
Its flexible meaning allows it to serve as a unifying theoretical basis for human rights and is part of the reason it
appears in the Universal Declaration of Human Rights, which states that ``all human beings are born free and equal in dignity and rights''  \citep{united1999universal}.
Although there are multiple notions of dignity, 
we focus on two.

The first is the notion of \emph{inherent dignity}, as popularized by Kant, who states that all humans possess ``a dignity (an absolute inner worth) by which he exacts respect for himself from all other rational beings in the world'' and that this dignity cannot be substituted, exchanged, gained, or lost \citep{kant2017kant}. 
Inherent dignity is based on the belief 
that, by virtue of being human, individuals must be afforded a ``necessary respect'' by others and the state \citep{gewirth20191}.
\citet{kant1967grundlegung} also believed that individual autonomy and  self-determination are special to humans 
and therefore intrinsically tied to dignity. 
In practice, inherent dignity is associated with negative liberty---a freedom from interference by the state that is rooted in the idea that a 
``person's dignity is best respected or enabled when he can pursue his own ends in his own way''
\citep{rao2011}.

The second notion of dignity relevant to this piece is \emph{dignity as recognition}, 
which requires that there be ``esteem and respect for the particularity of each individual'' \citep{rao2011}.
It demands that an individual's uniqueness is recognized and respected. 
Recall that inherent dignity is rooted in the idea that all individuals possess an inner worth that is deserving of respect regardless of whether their dignity is recognized.
By contrast, under the concept of recognition dignity,
an individual ``can have dignity and a sense of self only through recognition by the broader society'' \citep{rao2011}.
That all individuals inherently possess dignity is a ``presumption of human
equality'' \citep{rao2011}.  
On the other hand, dignity as recognition requires 
``treatment that \emph{expresses} the equal worth of all individuals and their life choices''  despite their differences \citep{rao2011}. 
Rather than freedom from interference, 
recognition dignity is a positive concept in 
that the state must protect recognition dignity by enforcing respect between citizens and designing policies that actively acknowledge the equal worth of each individual (or group) in their uniqueness \citep{rao2011}.
In the past, recognition dignity has been invoked in claims against defamation and hate speech as well as the right to develop one's personality \citep{post1986social,keegstra,grundgesetz2020}.

The respect for an individual's uniqueness that is demanded by recognition dignity is closely related to the right to be an exception. 
In highlighting how the reliance of data-driven decisions on rules can inflict harm on exceptions, 
the right to be an exception ``formalizes a basic respect for individual human dignity in a political system that otherwise allocates costs and benefits on the basis of majority rule'' \citep{paradis2015}. 
In this way, it can be viewed as a mechanism for protecting the recognition dignity of individuals in high-stakes, data-driven decision contexts. 
Recognizing the dignity of decision-subjects does not require that decisions always tip in their favor. 
It simply requires a respect for dignity---an acknowledgment that when a decision can inflict significant harm on the subject, 
the decision should be based on a ``respectful deliberation''  that balances the subject's unique circumstances alongside other considerations \citep{harel2014law}.

One may then wonder why the right to be an exception is needed given 
the importance that both international and domestic law already place on dignity. 
Despite its similarity to the right to be an exception, 
a right to dignity in data-driven decisions is too abstract to be operational on its own. 
As thoroughly examined by \citet{glensy2011dignity}, \citet{rao2011}, and  \citet{omahony2012dignity},
the concept of dignity is so malleable that it can be invoked in many, often conflicting ways. 
As they discuss, 
claims based on dignity ``are most likely to succeed when coupled with an underlying deprivation of individual rights,'' 
especially in the U.S. where the constitutional structure emphasizes negative liberties, 
because the amount of respect one's dignity demands is highly subjective and context-dependent
\citep{rao2011}.
As such, although dignity lays the foundations for the right to be an exception, 
the latter re-examines and refines it for data-driven decisions contexts.

\subsection{Right to individualized sentencing}\label{sec:individualized_sentencing}

In the previous section, 
we observed that attention to exceptions can be grounded in the notion of dignity; in particular, recognition dignity, which demands that society recognize and respect the unique attributes that distinguish individuals. 
In this section, 
we turn to an example of a right that invokes dignity.
Using the \emph{right to individualized sentencing determinations} as a case study, we examine how exceptions arise and have been addressed.
While we focus here on sentencing, the right to individualized decisions arises in other domains as well (e.g., in employment~\citep{cra-update}).
As discussed in Section \ref{sec:right}, 
the right to be an exception to a data-driven rule  is related to but goes beyond the right to individualized data-driven decisions (that is, individualization is one but not the only component of the right to be an exception). 

In the 1970s,
mandatory minimum sentences became more commonplace in the U.S. 
as part of an effort to ``make sentencing procedures fairer and sentencing outcomes more predictable and consistent'' \citep{incarceration2014growth}.
This change had unintended consequences, 
including shifting the power in sentencing determinations from the judge or jury to the prosecution
who could, at times, leverage mandatory minimum sentences to overcharge and obtain easy pleas \citep{berry2019individualized}. 
During this time, 
many states adopted mandatory death penalties for felony convictions, 
but rather than improve the fairness and consistency of sentencing, 
mandatory death penalties reduced the ability of sentences to reflect the degree of \emph{mens rea} revealed during trial \citep{berry2019individualized,woodson1976-2,McGautha1971}. 
It also forced the hand of juries---many refused to ``convict murderers rather than subject them to automatic death sentences'' \citep{woodson1976-2}.

In 1978, 
the U.S. Supreme Court ruled in \emph{Lockett v. Ohio} that defendants in capital cases are entitled to ``individualized sentencing determinations'' \citep{Lockett1978} due to the seriousness and irrevocability of the death penalty \citep{Furman1972}. 
Specifically, the Court ruled that the Eighth Amendment prescribes a ``fundamental respect for humanity'' that  ``requires consideration of the character and record of the individual offender and the circumstances of the particular offense'' for a sentence as serious as the death penalty \citep{woodson1976-2}. 
In 2012, 
the Court extended this concept to juvenile life-without-parole sentences in \emph{Miller v. Alabama}, arguing that life-without-parole constitutes an especially serious sentence
and that juvenile offenders are ``constitutionally different from adults for purposes of sentencing'' because ``juveniles have diminished culpability and greater prospects for reform'' \citep{Miller2012}.
More recently, 
 \citet{berry2019individualized} has argued that 
individualized sentencing determinations should be broadened to all felony cases 
because felony convictions carry serious consequences. 
 \citet{berry2019individualized} explains that felony convictions result in 
``dehumanizing effects that extend far beyond release, including the loss of right to vote, government surveillance, loss of possession and use of a firearm, housing consequences, employment consequences, and public benefits'' and therefore that the consequences of a felony conviction can be viewed as the ``death of one's non-felony self.''

The push for individualized sentencing determinations reflects a belief that, when the risk of harm is particularly high, 
a decision-subject's unique circumstances and how the decision may affect them 
deserves careful consideration.
In  \citet{berry2019individualized}'s words, sentencing determinations that are particularly serious ``ought to include consideration of all relevant
aggravating and mitigating evidence, and not flow automatically from the type of crime committed.''
Moreover, in practice, mandatory sentencing statutes re-delegate sentencing discretion from the court to the prosecution,  and ``prosecutorial decision-making, by contrast, occurs in a black box of secrecy,'' which is exacerbated by the fact that ``prosecutors enjoy a complete lack of accountability'' in how they set the sentence for the offender \citep{berry2019individualized}. 

Two parallels can be drawn from Berry's analysis.
First, individualized sentencing 
echoes the idea that there are conditions under which it is not appropriate to blanketly apply rules.
Individualized sentencing is therefore a movement towards greater consideration of exceptions when  decisions have serious consequences. 
Second, rules can be misused. 
Without sufficient transparency and accountability,
rules not only fail on the exceptions, 
but they also fall short in their original goals of producing more fair procedures and more consistent, predictable outcomes.

\subsection{A Note on Related Works} \label{sec:note_related_works}

There are many existing works on data-driven technologies and their pitfalls. These works have covered enormous ground, highlighting issues that arise during the application of data-driven technologies and gaps in their governance. 
We build on this literature, but there are four factors that together make this piece distinct. 

First, many works (such as those examining disparate impact \citep{barocas2016big}) focus on group-based outcomes, 
e.g., discrimination based on a protected attribute. 
In contrast, 
the right to be an exception examines data-driven decisions through the lens of individual outcomes rather than group-based ones. In particular, we discuss how one can determine if a data-driven rule is appropriate for a specific decision-subject, 
similarly to \citet{lippert2011we}. 

Second, most existing works propose to improve outcomes by requiring that data-driven tools be ``fair,'' ``accurate,'' and ``reliable''  \citep{citron2014scored,wachter2019right}. 
We find that such criteria are important but do not capture the full picture when evaluating the suitability of a data-driven tool for a \emph{specific} decision context. Accuracy, for instance, is an average notion---high accuracy only implies good performance in an average sense. Similarly, reliability implies good performance in a repeated sense---that, if run many times, an algorithm would consistently perform well. In this piece, we offer an additional consideration. In addition to fairness, accuracy, and reliability, 
there is another desideratum: a decision-maker should not presume that the data-driven rule is suitable for an arbitrary decision-subject,
particularly when the stakes are high. 
Rather, the decision-maker should only apply a data-driven rule if they are sufficiently confident (as measured against the risk of harm) that it is indeed suitable, as detailed in Section \ref{sec:right}.

Third, 
there are several works that examine whether data-driven rules should be sufficiently individualized in order to be applied. 
That is, they investigate how individualization addresses the problem of statistical discrimination \citep{lippert2011we,wachter2021fairness}. 
In this piece, 
we build on this discussion and argue that individualization is one, but not the only, component of evaluating a data-driven rule's suitability. 
We maintain that one must also consider the data-driven rule's \emph{uncertainty},
a concept that is often overlooked but is core to the right to be an exception. 

Lastly, our hope is to provide a framework that can serve as a common language with which to discuss data-driven exceptions across disciplines. 
To this end, we also consider the technical aspects of data-driven exceptions, including their origins (showing that data-driven exceptions arise in more ways than existing works typically consider, making the problem less straightforward than commonly assumed) and the technical viability of the proposed solutions. 
We pay particular attention to the latter. For example, although open-sourcing data-driven tools may be useful, it is infeasible in many cases (e.g., due to trade secret law or that open-sourcing introduces vulnerabilities to adversarial attacks). 
To ensure technical viability, we distill the right to be an exception down to three concepts, described in Section \ref{sec:right}, that are also meaningful in machine learning.

\section{The Right to be an Exception}\label{sec:right}

Rights are legally enforceable claims, such as the right to free speech and right to a fair trial. In recent years, multiple rights have arisen in response to new technologies, including artificial intelligence. For example, the right to be forgotten arose to protect individuals whose personal information appears on the internet—a technology that, in a sense, never forgets \citep{rosen2011right}. 

In this piece, we argue for an individual’s right to be an exception to a data-driven decision rule. The right says that, a decision-maker cannot presume that a data-driven rule is suitable for a given decision-subject---they must be sufficiently confident (relative to the risk of harm) that the individual is not an exception. 
In other words, a data-driven decision-maker—whether a machine or machine-aided human—must make a decision that inflicts harm only if they have applied due care and due diligence in determining whether the data-driven rule is fit for the decision-subject in question.  The greater the risk of harm, the higher the bar.  

Society has, for the most part, developed standards for assessing whether a human has applied due care and due diligence in decision-making (cf. the right to individualized sentencing \citep{berry2019individualized}). After all, the law has been honed to work for human-driven decisions. How one would operationalize this concept in the data-driven context is, however, unclear.  In this piece, we propose that adapting this requirement for data-driven decisions can be achieved by considering three factors: \emph{individualization}, \emph{harm}, and \emph{uncertainty}. Via these three components, we provide a concrete framework through which a decision-maker can determine when a data-driven rule is appropriate or a decision-subject can determine whether to contest a data-driven decision.

Importantly, establishing that the right to be an exception has been violated is not the same as establishing that the decision rule has made an error (i.e., that the data-driven rule made a mistake on the decision-subject). A mistake occurs when a prediction does not match the observed outcome. In other words, the target outcome must be \emph{observed} in order for a rule to have made a mistake. 
The right to be an exception, on the other hand, can be violated even if the outcome of interest is not observed. 
For example, consider a judge deciding whether to deny bail, which is largely based on whether the judge believes the defendant will fail to appear for their court date. Suppose a defendant is denied bail. Then, they cannot prove that they would have appeared \emph{if} they had been granted bail instead. It is therefore impossible to determine if the decision to deny parole was a mistake. One can, however, assess whether the decision to deny parole was justified under the circumstances, i.e., whether the decision respected an individual's right to be an exception to a data-driven rule. In this piece, we propose that such an assessment is possible for data-driven decision rules through the consideration of individualization, uncertainty, and harm.  

\subsection{Individualization: Moving from the Aggregate to the Individual}

For many, the natural first step to designing a data-driven rule that surpasses the appropriate levels of care and diligence in ruling out an exception is individualization: the process of tailoring a rule to the specific circumstances under consideration. In short, individualization shifts attention from the aggregate to the individual. The more individualized a rule, the more suitable it is for a particular decision-subject. For example, one way to make a data-driven rule more individualized is to add features, or inputs, to the model. A data-driven rule that uses an applicant’s age, home address, and occupation in order to decide whether to grant a loan is therefore more individualized than one that uses only their age and home address.

Individualization is an information concept in that it requires a decision-maker to consider the totality of an individual’s circumstances rather than make judgments based on a limited set of information. In other words, to individualize a rule is to give it additional (relevant) information. The desire for individualized decisions—the first component of the right to be a data-driven exception—is not new. Indeed, \citet{lippert2011we} discusses the right to be treated as an individual as a proposal for reducing statistical discrimination (treating an individual as if they were the statistical average of similar individuals). The push for individualization is based on the logic that, the more individualized an assessment, the less likely it is to have made broad-strokes generalizations and, as a result, to yield exceptions. 

Individualization is a particularly useful concept because it appears in both legal texts (cf. the right to individualized sentencing \citep{berry2019individualized,jorgensen_2021}) as well as technical ones. As such, a law requiring individualization in data-driven rules would pave a clear path for computer scientists. Indeed, much of machine learning echoes the belief that, with enough information and enough historical data, a data-driven rule can predict the target outcome with perfect accuracy. Individualization has become so central to machine learning that data-driven rules are often justified based on their level of individualization. Most theorems in machine learning, for instance, follow the template: “As N goes to infinity, the error goes to 0” (occasionally accompanied by a “with high probability”), where N quantifies the amount of information.

Perfect individualization, however, is difficult to implement. In practice, current methods are incapable of individualizing in ways that humans do naturally. Humans, for example, are generally flexible enough to update their decisions to incorporate additional pieces of information. Although a judge may initially receive certain information about a defendant, they can update their belief when given novel information (e.g., that the defendant volunteers or has dependents). Humans rely on this unique ability to holistically examine an individual’s circumstance in order to produce individualized decisions. In contrast, most (if not all) data-driven rules have fixed inputs and cannot incorporate features that are not present in the training data.

So, perhaps perfect individualization is not possible, but is individualizing the rule as much as possible (albeit imperfectly) all that is required to ensure that the rule is fit for use? Stated differently, suppose that a data-driven rule were perfectly individualized—that is, it incorporates all relevant information. Would such a fully individualized data-driven rule uphold an individual’s right to be an exception? 

\subsection{Individualization is Not Enough: Uncertainty Also Matters} 

No—individualization is not the only pertinent factor. There are two additional components: uncertainty and harm, and we focus on the former in this section. The takeaway is that while individualizing a data-driven rule takes an important step toward ensuring that it does not neglect relevant information, no amount of individualization can remove all the uncertainty in a data-driven rule, and the amount of uncertainty matters when the risk of harm is high. 

Recall that individualization is an information concept: it relies on the belief that, holding everything else equal, adding information improves a data-driven rule. Conveniently, this reasoning also underlies machine learning, which is founded on the idea that data is king (i.e., that with enough information, a data-driven rule can perform perfectly). In reality, however, even the best data-driven models make mistakes, often because some predictions are inherently impossible to get right every time. In fact, there are very few (if any) meaningful settings in which a perfect rule exists, and the main barrier is uncertainty. 

To illustrate the limitations of individualization, consider the following two types of uncertainty \citep{kendall2017uncertainties}: 
\begin{enumerate}
	\item \emph{Epistemic uncertainty} is systematic or reducible uncertainty that arises from lack of knowledge. For example, a prediction of tomorrow’s temperature that is based on past years’ temperatures at this time of year has greater epistemic uncertainty than the prediction of tomorrow’s temperature based on past years’ temperature at this time of year \emph{and} today’s temperature. 
	
	\item \emph{Aleatoric uncertainty} is statistical or irreducible uncertainty that arises from the inherent randomness or “unknowability” of an event. At the time of prediction, no information exists that can reduce this type of uncertainty. For example, the randomness in the wind patterns that may occur between today and tomorrow prevents a temperature prediction that is made today from being perfectly certain about tomorrow’s temperature, and this randomness can be attributed to aleatoric uncertainty. 
\end{enumerate}
Through these two types of uncertainty, it becomes clear that while individualization may reduce epistemic uncertainty, it cannot reduce aleatoric uncertainty. In some cases, individualization does not even reduce epistemic uncertainty. Consider the following examples. 

\begin{example}[Individualization increases granularity at the risk of increasing uncertainty]
	Consider a data scientist who wishes to increase the individualization of a data-driven rule used in healthcare. To do so, the data scientist adds features to the rule’s input. Instead of taking in a patient’s current age, height, and weight as inputs, the data scientist modifies the rule to also accept the patient’s history of heights and weights at every year of their life. 
	
	Suppose the data scientist uses a nearest-neighbors-style algorithm—an approach that makes a prediction for patient X based on previous (exemplar) patients who have similar attributes as X. Then, the more refined the features, the fewer exemplar patients for X exist. In other words, individualization reduces the amount of evidence that the nearest-neighbors rule can use to generate its assessment. As such, while the data scientist reduces epistemic uncertainty in one way, they increase it in another. 
\end{example}

\begin{example}[The unknowability of unobserved outcomes]
	Consider a data-driven decision aid for college admissions—specifically, one that predicts how well a student will perform if admitted. Beyond random noise, there are multiple ways that aleatoric uncertainty arises. 
	
	For one, even if the student is similar to previous students for which there is data, one could argue that a student's performance is \emph{not predetermined}, i.e., that they have the ability to perform differently from past individuals. 
	That each student possesses their own potential for success---that they have their own autonomy---means that no amount of individualization can predict performance with certainty. 
	Indeed, believing that a data-driven rule carries no uncertainty  holds students responsible for the performance of previous students (namely, students in the training data). 
	While individualization can improve a data-driven prediction, 
	it continues to hold the decision-subject responsible for the performance of previous---albeit, increasingly similar---students, 
	and there is always uncertainty associated with the decision-subject's own potential for success. 
	
	For another (and perhaps more concretely), there is also omission bias. 
	The training data only captures the performance of students who were admitted, which implies that the performance of a student who was not admitted is unknowable \citep{kleinberg2017}. Perhaps a student who is similar to the decision-subject but was not admitted would have performed very well. 
	
	Lastly, even if a decision-maker has perfect knowledge of previous students' outcomes, any decision that is made now can only use information obtained up until this moment. There are, however, countless factors (or, in the language of causal inference, “interventions”) that could influence a student’s performance between the time of acceptance and graduation, such as whether they receive tutoring, who they befriend, and whether they take a part-time job. The only way that an assessment can be perfect and rid of uncertainty is for the target outcome itself to be an input to the assessment, but this logic is circular. If one could measure the target outcome, one would not need to infer it.
\end{example}

\subsection{The Importance of Uncertainty: Weighing the Risk of Harm}

In short, individualization can, at best, remove epistemic uncertainty, but no amount of individualization can remove aleatoric uncertainty. Perhaps one of the best ways to summarize this argument is via \emph{computational irreducibility} \citep{wolfram2002new}. The reasoning behind this concept goes: a computer is one of many components in our world. Therefore, the complexity of a computer must be strictly lower than the complexity of the world. It follows from this logic a computer cannot predict any arbitrary outcome of interest $Z$ (even if it was given all the historical data in the world and continually fed new data) because the complexity of the process that produces $Z$ may be higher than the computational capacity of the computer. 

That is not to say that data scientists should throw up their hands and give up. Indeed, computational irreducibility does not imply that every prediction task is hopeless. Rather, it says that uncertainty is inevitable when predicting a \emph{complex} target outcome. However, eliminating uncertainty is besides the point. It is unreasonable to ask for a perfect data-driven rule that makes no mistakes. Instead, the right to be an exception asks that the level of uncertainty be \emph{balanced against the risk of harm}. 

More precisely, suppose that one of the decision outcomes would inflict significant harm. Then, no matter how individualized a decision rule may be, the decision to inflict harm should only follow if the level of certainty is high enough. If, on the other hand, the level of uncertainty (epistemic and aleatoric) is too high, then the decision-maker should err on the side of caution (less harm).

As an extreme example, suppose a decision-maker is presented with a newborn and must decide whether to confine them for the rest of their lives based on an evaluation of whether they will commit murder during their lifetime. The decision is made at the time of birth, so the only information that is available must also be available at the time of birth. A rule could be perfectly individualized (based on the information at the time of birth), but most would agree that there are so many unknowable factors that could contribute to the newborn's future actions that no amount of individualization would justify inflicting a harm as high as confining a newborn for life. 

A decision outcome’s risk of harm therefore determines the amount of individualization and certainty necessary to utilize a data-driven rule whose recommendation inflicts harm. 
Some decisions might carry a risk of harm so low the level of individualization and certainty needed to justify the use of a data-driven rule is accordingly low. It is natural to then ask: How should harm be measured? While providing an explicit framework for quantifying harm is out of the scope of this piece, we note that prior works have laid out a path for doing so, including  \citet{wachter2019right}'s work on the right to reasonable inferences (in which they discuss the determination of “high-risk inferences”) and \citet{kaminski2021right}'s right to contest AI (in which they characterize risk of harm in terms of “significant effects”). The European Union's Artificial Intelligence Act also provides a “risk methodology” for categorizing high-risk decision contexts \citep{euAIact}.

\subsection{Putting it All Together}

The right to be a data-driven exception requires that the decision-maker not necessarily presume that a data-driven rule is suitable for a decision-subject, particularly when the risk of harm is high. 
Rather, the right requires that the decision-maker
inflict harm only if they have applied due care and due diligence in determining whether the data-driven rule is fit for use on the individual in question. It seeks to prevent data-driven decisions from inflicting irreparable and repeated harm on individuals who, through no fault of their own, are exceptions to a data-driven rule. This right emphasizes that data-driven rules cannot be applied blanketly. While a data-driven rule may be appropriate for some individuals, it may not be appropriate for all. In particular, when a decision may inflict significant harm on the decision-subject, examining whether or not a decision-maker is justified in using the data-driven recommendation becomes pertinent. In this way, the right is in keeping with existing concepts (originally intended for human decision-makers), including the right to dignity and the right to individualized sentencing. 

Importantly, the right does not imply that data-driven rules should be dropped altogether, nor does it suggest that they be used in every case. The right does not even suggest that there is a clear line between the types of decisions in which data-driven rules are appropriate (e.g., that data-driven decision aids should be used in lending but not sentencing). Rather, the right argues that there are some contexts in which the stakes are so high that each decision-subject deserves appropriate consideration of whether the data-driven rule is fit for them. In the same way that certain information is discarded as irrelevant (e.g., a college admissions board may discard a student’s sophomore Fall grades if a family tragedy occurred that semester), a data-driven recommendation may need to be discarded. While useful, this analogy does not carry over perfectly because it is unclear when to discard a data-driven rule. Data-driven rules behave quite differently from human ones—for instance, the “intent” and “reasoning” behind a data-driven recommendation are often inscrutable. 

In this piece, we find that the right to be an exception requires the consideration of three factors: individualization, uncertainty, and harm. Crucially, these three factors are not only interpretable to lawmakers, but also meaningful concepts in machine learning. They therefore provide a clear language with which to assess data-driven decisions.  

More precisely, the right to be a data-driven exception requires that the decision-maker first evaluate the level of harm of each decision outcome. Based on the level of harm, the decision-maker should then evaluate the data-driven rule based on two considerations: individualization and uncertainty. Individualization characterizes the suitability of a rule based on how much information it considers (e.g., whether it knows enough about the decision-subject or has enough training data that pertains to the decision-subject). The level of uncertainty can be divided into two types: epistemic and aleatoric. The former captures uncertainty due to lack of information, and the latter captures the inherent unknowability of a prediction task. The right requires that the decision-maker utilize the data-driven recommendation only if the levels of individualization and certainty are high enough to justify the level of harm that would result from that recommendation.

\section{Operationalizing the Right to be an Exception}\label{sec:operationalizing}

In this section, we examine how the right to be an exception could be operationalized. 
We consider what it does (and does not) mean to invoke the right as well as \emph{ex ante} and \emph{ex post} measures.

\subsection{Invoking the Right}\label{sec:invoking}

Does invoking the right to be an exception to a data-driven rule simply mean proving that the data-driven rule made a mistake, making it similar to the right to rectification \citep{gdpr}?
Or is it that a decision-subject who does not like their decision outcome can always claim to be an exception, thus nullifying any data-driven rule in high-risk settings?

The right to be an exception says neither. 
Claiming the right to be an exception to a data-driven rule is \emph{not} equivalent to proving that the data-driven rule made a mistake. 
For one, 
the outcome of interest is not always observable. 
In many cases, it is impossible to determine whether a mistake was made (e.g., a judge can never know whether a defendant who is denied parole would have reoffended \emph{if} they had been granted parole instead). 
As such, a right to rectification would require that the outcome is observed, 
but the right to be an exception does not. 
That is, 
\emph{the right to be an exception can be violated without a mistake occurring}. 

Consider the following (simplified) example. 
Suppose a data-driven rule delivers random recommendations. 
For instance, suppose that it simply flips a coin each time it is asked for a recommendation. 
Even this random rule is bound to be correct for some individuals. 
However, whether this rule happens to be correct is besides the point. 
If the decision's risk of harm is high (e.g., a sentencing decision), 
such a rule should not be applied regardless of whether or not it turns out that, 
down the line, 
the random flip happens to correctly predict the outcome. 
It is simply not suitable for a high-risk setting. 
This evaluation of a data-driven rule's suitability is what underlies the right to be an exception. 
Namely, the data-driven rule should only be applied if deemed suitable for the specific decision-subject, where the level of consideration must be fitting to the risk of harm (for which we provide a framework in Section \ref{sec:right}).

Importantly (and in answer to the second question above), 
the right to be an exception does not imply that \emph{every} individual is an exception. 
That is, a decision-subject who does not like their data-driven decision outcome cannot simply claim that the right to be an exception nullifies the data-driven decision. 
In fact, 
\emph{a data-driven rule can still uphold the right to be an exception even if it makes mistakes}. 
It is indeed unreasonable to expect a data-driven rule to never make mistakes---a decision-subject can, at best, hope that a data-driven decision-maker ensures that a data-driven recommendation is only used if it is deemed fit for the given context. 
The right to be an exception captures this principle. It does not requires that a data-driven rule is perfect but that is appropriately applied. 

In this way, 
the right to be an exception is not a matter of mistakes. 
It can be violated even when a mistake has not been made (or cannot be verified). 
At the same time, 
the right is not necessarily violated when a mistake is made.
Crucially, 
while a data-driven rule's accuracy---which many believe can be used to evaluate a data-driven rule's suitability \citep{loomis2016}---is an important performance metric, 
it is another way of measuring mistakes. 
Therefore, accuracy alone cannot fully capture the suitability of a data-driven rule, as detailed in Section \ref{sec:right}.

\subsection{\emph{Ex ante} Justification}\label{sec:ex_ante}

The right to be an exception 
would require an  \emph{ex ante} justification that a data-driven decision appropriately considers the three main components of the right---harm, individualization, and uncertainty---before such a data-driven decision is applied. 
Specifically, 
the data-driven assessment must 
(1) evaluate the potential harm that the decision could inflict; 
(2) justify the rule on the basis of its level of individualization;
and
(3) demonstrate that, given the level of harm and individualization, the rule appropriately and {meaningfully} incorporates uncertainty into its decision \emph{or} appropriately and {meaningfully} communicates it to the final decision-maker. 

In order to evaluate a decision's potential harm, one can use the standard of ``significant effects'' in Article 22(1) of the General Data Protection Regulation (GDPR) \citep{gdpr}, 
as studied by \citet{kaminski2021right}. 
\citet{wachter2019right} present a similar framework in their discussion of ``high-risk inferences'' with respect to the right to reasonable inferences. 
One can also turn to the  \citet{euAIact}'s Artificial Intelligence Act, which provides a ``risk methodology''
for evaluating and categorizing high-risk decision contexts. 

If the risk of harm is high enough, 
the next step is to characterize a data-driven rule's level of individualization (which may vary across decision-subjects), as given by (2). 
Characterizing a rule's level of individualization can be done in multiple ways. 
For example, one could require that a data-driven decision aid report its input variables, which reflect the data-driven rule's granularity. 
One could also require that the data-driven rule be evaluated on performance metrics that are more fine-grained than accuracy, 
such as calibration or multicalibration \citep{hebert2018multicalibration}.

Lastly, 
(3) is a final step that combines the insights of (1) and (2). 
Specifically, (3) determines whether, 
given the potential harm assessed in (1) and the level of individualization found in (2), 
the final decision appropriately and meaningfully incorporates uncertainty. 
If the final decision maker is the \emph{algorithm}, 
one must demonstrate that the assessment appropriately and meaningfully considers uncertainty. 
If the final decision-maker is \emph{human}, then the data-driven assessment must appropriately and meaningfully communicate uncertainty to them. 
Incorporating uncertainty is necessary, as it synthesizes the assessments of harm and individualization.
For instance, 
if a decision carries a risk of significant harm, 
then the level of individualization and the accompanying certainty may not be high enough to justify inflicting harm. 
It may even be the case that, in certain contexts,
no matter how individualized the assessment, 
there is too much uncertainty to justify inflicting harm while, in others, the risk of harm is so low that a high level of uncertainty is acceptable. 
Meaningfully incorporating or communicating uncertainty for (3) is an active area of research in human-computer interaction \citep{hullman2016evaluating,hofman2020visualizing}. 
To communicate uncertainty meaningfully, 
the assessment could report on different types of uncertainty, 
similarly to how existing works distinguish between epistemic and aleatoric uncertainty \citep{NIPS2017_2650d608}, 
as discussed in Section \ref{sec:right}.

\subsection{\emph{Ex post} Contestation}

It is important that decision-subjects be able to contest a data-driven decision \emph{ex post} by claiming the right to be an exception. 
As explained in Section \ref{sec:contest_legitimacy} and explored by \citet{kaminski2021right}, 
contestation is an accountability mechanism that enhances the legitimacy of data-driven assessments as well as builds the public's trust in them. 

As a possible template, one could turn to the procedure for contesting on the basis of Title VII of the U.S. Civil Rights Act's notion of disparate impact \citep{barocas2016big}. 
Specifically, 
Title VII prohibits employment discrimination due to the individual's race, color, religion, sex, or national origin.
A plaintiff---an individual who believes that their (potential) employer violated Title VII---can sue the employer by providing evidence of what is known as ``disparate impact.''
 
In disparate impact cases, 
a plaintiff must first establish that an employment practice causes negatively impacts a class of individuals protected by Title VII compared to its impact on individuals outside the protected class. 
Even if disparate impact is established, however, 
it can be countered if the defendant---or employer---successfully shows that the employment practice is rooted in ``business necessity.'' 
The defense of ``business necessity'' can further be refuted if the plaintiff provides a compelling alternate employment practice that would mitigate disparate impact without violating business necessity. 
Contestation on the basis of the right to be an exception could mirror this three-stage procedure, as follows.
First, the plaintiff must establish that (1), (2), and/or (3) from Section \ref{sec:ex_ante} has been violated by the data-driven decision. If the plaintiff is successful, 
the defendant can counter by showing that the data-driven decision could not have been changed without demanding significant resources or inflicting disproportionate harm on other parties.
Finally, if the defendant is successful in this second stage, 
the plaintiff can refute the defendant's justification by providing an alternate procedure that improves upon the assessment with respect to (1)-(3) and does not demand excessive resources or inflict disproportionate harm on other parties. 
This procedure is one among many contestation mechanisms, as surveyed by \citet{kaminski2021right}.

\section{The Gap the Right to be an Exception Fills}\label{sec:gap}

The right to be an exception fills an important gap in the governance of data-driven decisions. 
As discussed in Section \ref{sec:dignity}, 
although dignity underlies the right to be an exception, 
dignity is too malleable a concept for our purposes. 
Furthermore,
the unique challenges that arise in data-driven decision-making
create opportunities for significant and widespread harm (see Section \ref{sec:data_driven_exceptions}).
As stated by \citet{kaminski2021right},
``[artificial intelligence] warrants unique risks that deserve distinct treatment.''
In this section, we unpack the ways the right to be an exception fills a gap in the governance of data-driven decisions.

\subsection{Legal Recourse Lends Legitimacy}\label{sec:contest_legitimacy}
Establishing the right to be an exception provides individuals with a path to {contest} data-driven decisions that blanketly apply data-driven rules.
The ability to contest data-driven systems is highly important. 
Instead of trusting on faith alone that data-driven rule are suitable for every context on which they are applied, 
the ability to contest gives individuals {agency} over the decisions that affect their lives \citep{kaminski2021right}.  
In doing so, ``a fair contestation process can enhance the perceived {legitimacy}'' of and {trust} in data-driven decisions,
to the benefit of both decision-subjects and makers \citep{kaminski2021right}. 
On one side, algorithm designers can continue to develop and deploy. 
On the other side, decision-subjects can serve as checks, 
ensuring that the algorithms are applied appropriately by helping to identify when they fail.
In this way, contestation serves as  a ``systematic management technique'' that smooths the integration machine learning into decision-making by ``uncovering errors, identifying their causes, and providing schemes and incentives to correct them''  \citep{crawford2014big}.

Affording individuals the ability to contest data-driven decisions on the basis of the right to be an exception does not mean that data-driven rules should be abandoned. 
The right serves as a legal complement to efforts by the computer science community to improve performance on all cases, including the exceptions. 
Rather than bar data-driven rules from decision-making, 
contestation identifies directions for algorithm improvement,  ``provid[ing] individuals recourse even when they choose to continue to participate in the activity”
\citep{kaminski2021right}.

A fair contestation process also articulates the level of \emph{transparency} an algorithm should exhibit. 
In the absence of perfect {transparency}---a concept that continues to elude us because algorithms, even with full access, can be difficult to intuit---contestation serves as a rigorous alternative that sets guidelines for how transparent a system must be, 
which is to say as transparent as needed to determine whether it upholds an individual's rights.

\subsection{Shifting the Burden of Proof} 
By shifting power away from decision-makers who are currently permitted to argue that excellent (or even good) average-case performance justifies the poor treatment of the few exceptions, the right to be an exception also rebalances the burden of proof. 

At its core, the right to be an exception expresses that the presumption should not be that a data-driven rule is fit for use on any arbitrary decision-subject, especially when the risk of harm is high. 
Instead, one must presume that the data-driven rule is not fit, only applying it when sufficiently convinced that the rule's levels of individualization and certainty justify the level of harm it may inflict. 
In this way, the right shifts the burden of proof toward the data-driven decision-maker. 
At the moment, plaintiffs ``face an uphill battle ... with regards to big data inferences''  \citep{kaminski2021right,ajunwa2021auditing,barocas2016big}. 
Unlike with humans, 
it is often difficult to gain insights into an algorithm's assessment
and one cannot base a legal claim on the algorithm's ``intent'' (e.g., racial animus). 
In place of these normal routes, 
algorithms are often justified using evidence of good performance,
such as high accuracy. 
However, as noted in this piece, 
most common metrics, including accuracy, are {averages} that do not indicate an rule's suitability for an arbitrary decision-subject,
making them unfit in high-stakes, non-repeatable contexts. 
Therefore, when individuals contest data-driven decisions, 
they must argue that an algorithm is unsuitable on the basis of averages, which tends to fall in favor of the decision-maker. 
The right to be an exception redistributes this burden, 
requiring that decisions at risk of inflicting significant harm demonstrate adequate attention to individualization, harm, and uncertainty.

As an example, 
consider 
\emph{State v. Loomis} \citep{loomis2016}. 
In 2013, Eric Loomis was charged in relation to a drive-by shooting. 
Although Loomis denied participating in the shooting, he conceded to driving the same car that day and pleaded guilty to “attempting to flee a traffic officer and operating a motor vehicle without the owner’s consent,'' but not to the shooting \citep{loomis2016}. 
An algorithmic risk assessment, whose methodology is a trade secret, was consulted as a part of his sentencing determination, 
and Loomis was sentenced to six years of imprisonment and five years of extended supervision.
Loomis filed for post-conviction relief. 
He argued (a) that the use of an algorithmic risk assessment  infringed on both his right to an individualized sentence and his right to be sentenced on accurate information; 
and (b) that the risk assessment---and thereby the court---unconstitutionally used gender to determine his sentence. 

Loomis was denied post-conviction relief on two grounds.
First, that the use of gender ``served the nondiscriminatory purpose'' of  improving the algorithm's accuracy \citep{loomis2016}. 
Second, because the risk assessment uses only publicly available data and information about the defendant, Loomis ``could have denied or explained any information that went into making the report and therefore could have verified the accuracy of the information used in sentencing'' \citep{loomis2016}.
We contend that the court made several missteps in their ruling. 
For one, by claiming that gender serves a ``nondiscriminatory purpose,''
the court appealed to the concept of intent, 
a concept lacks meaning when assigned to algorithms. 
For another, using accuracy to justify the algorithm prevented Loomis from receiving more serious consideration about whether an \emph{aggregate} measure like accuracy was suitable for his \emph{specific} case. 
Finally, arguing that Loomis could have ``denied or explained'' the information used by the data-driven assessment placed an enormous burden of proof on the plaintiff who would need to verify the training data.

Although the court recommended that judges should be provided with warnings of the shortcomings of risk scores, 
some have argued that, unless algorithm designers are expected to provide more meaningful information about the assessments,
``judges will not be able to calibrate their interpretations'' and cannot know ``how much to discount these assessments'' \citep{Wisconsin2017}.
To this end,
the right to be an exception requires the uncertainty of a data-driven rule be meaningfully incorporate or communicated in the decision-making process. 
In this way,
the right to be an exception shifts the large burden of proof currently placed on decision-subjects toward decision-makers, 
who would be compelled to demonstrate that a data-driven rule appropriately balances individualization, harm, and uncertainty before applying it.

\subsection{Unraveling the Black-Box}
We conclude with two final notes. 
The first is that, due to the scale at which data-driven rules are being applied, 
the right to be an exception may serve slow data-driven {feedback loops}. 
These feedback loops occur because data-driven rules be applied with great efficiency, often to the detriment of a subset of the population. 
For example, suppose that a welfare algorithm performs well on the large majority but poorly on a small number of individuals whom it mistakenly classifies as high-risk for welfare fraud. 
If this algorithm is used by many welfare agencies, 
then the same individuals are consistently denied welfare.
Because the individuals that an agency chooses to grant welfare are eventually used to train future algorithms, 
these individuals will continue to be denied welfare,
leading to a feedback loop. 
The right to be an exception can slow such feedback loops by compelling decision-makers to pay greater attention to the exceptions. 
Notably, the right to be an exception is not subsumed by discrimination law because, due to the complex and unintuitive nature of data-driven rules, data-driven exceptions do not fall neatly along demographic lines. 

As our second and final note, the uncertainty component of the right to be an exception may mitigate challenges posed by the inability to access or intuit algorithmic logic. 
When humans provide assessments, 
a decision-maker is often able to detect intent (e.g., racial animus) and places a degree of trust in the assessment accordingly. 
Although algorithms lack intent, 
one of the primary purposes of detecting intent is to assign credibility to an assessment. 
In the absence of intent, 
the right to be an exception encourages the decision-maker to incorporate assessments with appropriate caution, 
effectively replacing a credibility assignment. 

\section{Conclusion}

It is widely acknowledged that the governance of data-driven decisions requires new concepts and tools. 
In this work, 
we argue that decision-subjects have the right to be an exception to a data-driven rule. 
That is, 
the presumption should not be that a data-driven rule---even one that has high accuracy---is suitable for an arbitrary decision-subject of interest. 
Rather, a decision-maker should only apply a data-driven rule if they have applied due care and due diligence (relative to the risk of harm) in excluding the possibility that the decision-subject is an exception to the given data-driven rule. 
In some cases, the risk of harm may be so low that only cursory consideration is required. 
In others, the risk of harm may be so high that a decision-maker must be convinced that the data-driven rule works well on the \emph{specific} decision-subject of interest before applying it. 

To claim the right in practice, 
we provide a three-part framework for determining whether a data-driven decision respects an individual's right to be an exception to a data-driven rule. 
The framework requires that a decision-maker evaluate the data-driven rule based on its \emph{individualization}, \emph{uncertainty}, and \emph{harm}. 
We unpack these three components in detail. 
Importantly, 
we find that evaluating a data-driven rule on the basis of accuracy or individualization alone is not enough to justify the use of a data-driven rule when the risk of harm is high. 
Accuracy reflects a data-driven rule's average performance---it does not speak to a rule's performance on a particular decision-subject---and must therefore be considered alongside other factors. 
At best, individualization reduces epistemic uncertainty,
but it alone cannot characterize the suitability of a data-driven rule for a given decision-subject, 
particularly when the decision context is complex. 
The right to be an exception requires looking beyond accuracy and individualized decision-making. 
As we show in Section \ref{sec:right},  
the right to be an exception involves the balance of three factors---individualization, uncertainty, and harm. 
Because these three factors are also meaningful concepts in machine learning, 
this framework provides a common language with which to assess data-driven rules. 

Claiming the right to be an exception does not mean proving that a mistake has been made, nor does it mean that a rule that makes mistakes does not uphold the right. 
We explore the implications of the right to be an exception as well as \emph{ex ante} justifications and \emph{ex post} mechanisms that should accompany the right in Section \ref{sec:operationalizing}.

\bibliographystyle{chicago}
\bibliography{ref.bib}

\end{document}